# Laboratory grazing-incidence X-ray fluorescence spectroscopy as an analytical tool for the investigation of sub-nanometer CrSc multilayer water window optics

Veronika Szwedowski-Rammert[a], Philipp Hönicke[b], Meiyi Wu[c], Ulrich Waldschläger[d], Armin Gross[d], Jonas Baumann[a], Gesa Goetzke[a], Franck Delmotte[c], Evgueni Meltchakov[c], Birgit Kanngießer[a], Philippe Jonnard[e], Ioanna Mantouvalou[a*]

[a]Institute for Optics and atomic Physics, Technical University of Berlin, 10623, Berlin, Germany
[b]Physikalisch-Technische Bundesanstalt, Abbestr. 2-12, 10587 Berlin, Germany
[c]Institut d'Optique Graduate School - Paris Saclay, 2 avenue Augustin Fresnel 91127 Palaiseau cedex, Paris, France
[e]Bruker Nano GmbH, Berlin, Germany
[e]Sorbonne Université, Faculté des Sciences et Ingéniérie, UMR CNRS, Laboratoire de Chimie Physique-Matière et Rayonnement, 4 place Jussieu, F-75252 Paris cedex 05, France
*Corresponding author: ioanna.mantouvalou@helmholtz-berlin.de
Current affiliation: Helmholtz Zentrum Berlin, Germany

**Abstract**
Efficient multilayer optics for radiation in the water window range are difficult to manufacture due to extremely small layer thicknesses and severe intermixing of elements between the layers. Therefore, adequate analytics and short feedback loops are of utmost importance for manufacturers to improve performance and efficiency. We show the possibility for non-destructive elemental depth profiling with commercial laboratory equipment using four real-life CrSc multilayer samples. Comparative measurements at the laboratory of PTB at the synchrotron radiation facility BESSY II confirm the results and prove the potential of laboratory equipment for the reliable analysis of stratified materials with sub-nanometer layer thicknesses.

## 1. Introduction

Optical systems often require redirection of photon beams which becomes increasingly more complicated for small wavelengths as the refractive index becomes smaller than unity. Periodic multilayer (ML) optics rely on the principle of constructive interference due to reflection on interfaces and can reach efficiencies of tens of percent for specific wavelengths (such as EUV). Depending on the material selection and structure design of the multilayers, these mirrors have their peak reflectivity at specific wavelengths.

The CrSc based ML system is a promising candidate as high reflective mirror in the soft X-ray range and particularly in the water window (WW, between oxygen and carbon K absorption edge, from 2.3 to 4.4 nm) [1]. For radiation in this spectral range, the absorbance of biological material, of which carbon is the main component, is more than one order of magnitude larger than the usually surrounding water. This enables observing living specimen in aqueous environment and, thus, contributes to the development of scientific instruments such as microscopes dedicated to biological samples [2-4].

The combination of alternating sub-nanometer Cr and Sc layers yields a high contrast in refractive index above the Sc L edge and results in a calculated reflectivity of 60 % at 3.11 nm [5]. In practice however, the ML optics only reach 32 % due to well-known diffusion which directly reduces the optical contrast at the interfaces and consequently limits the optical performance of the mirror [6].

A fine understanding of the substructure of a manufactured optic and the involved processes between the materials constituting the stratified structures is thus crucial for developing, optimizing and eventually improving the performance of the multilayer. Therefore, increased effort is dedicated to their characterizations, preferably in the laboratory. To circumvent intermixing of layers diffusion barriers can be introduced which prevent interdiffusion. In this work CrSc ML samples with an additional $B_4C$ layer are investigated. The samples are deposited on sliced and polished Si (100) wafers using magnetron sputtering, see [7, 8] for details.

Due to the sub-nanometer thin individual layers and the strong intermixing processes characterization of the samples is challenging. As has been shown by Haase *et al.* [9], several complementary methods are necessary to find a consistent sample model with low uncertainties for all evaluated parameters. One possible nondestructive method which can be used to characterize the in-depth elemental composition of CrSc multilayers is grazing incidence X-ray fluorescence (GIXRF) [10]. This analytical technique employs X-ray radiation to generate X-ray fluorescence from the sample. This fluorescence is element specific, thus the reconstruction of the elemental composition of individual layers of the sample is possible. A depth distribution of the elemental composition within the sample is concluded from measurements under distinct incidence angles; each angle corresponding to a specific information depth. The result of a GIXRF measurement is an angular profile where the intensities for each relevant fluorescence line obtained from a deconvolution process are plotted over the angular position at which the corresponding spectrum was measured. When the incident X-ray radiation on the sample has high temporal and spatial coherence, i.e. high monochromaticity and low divergence, an X-ray standing wave (XSW) field is generated by interference of the incident beam with the reflected beam [11,12]. Strong changes in the refractive index, i.e. sharp elemental boundaries, lead to increased reflection compared to smooth transitions, and thus a more pronounced XSW field. As a result, a characteristic interference pattern can be observed in the angular profile. The shape of the interference pattern is predominantly influenced by the elemental depth distribution and the quality of the interfaces of the individual layers of the sample; a rough or diffuse interface lowers the contrast of the interference while the thickness of the layers shifts its angular position. To quantify the measurements the data are compared to simulated GIXRF profiles from sample models. The simulations are adapted to fit the measured data by varying initial sample parameters (e.g. layer thickness, density, roughness).

GIXRF is often used at synchrotron radiation facilities, due to high available photon fluxes, low beam divergences and the demand for a reliable knowledge about geometry parameters which lower the systematic uncertainty of the results. As the access to beamtime at synchrotron radiation facilities is limited and direct laboratory access is desirable especially for materials development, laboratory setups for angular resolved XRF investigations are increasingly developed and tested [13-15]. One major advantage of GIXRF is that no to minimal sample preparation and particularly no sectioning is necessary for the measurements. This fact and the robustness of X-ray tube sources render the development of reliable laboratory setups feasible which can be used on-site at manufacturing facilities and ultimately for process control feedback loops.

We present measurements on four CrSc multilayer samples with laboratory equipment and compare the results with results obtained from experiments using synchrotron radiation at the Physikalisch Technische Bundesanstalt (PTB).

## 2. Measurements

The laboratory measurements are performed with a Bruker S4 T-STAR™. This instrument was designed for total reflection XRF (TXRF) applications and upgraded with the option to perform angular scans and thus GIXRF measurements [16]. The instrument is equipped with a 50 W Mo microfocus X-ray tube (50 kV, 1 mA) and a 60 mm² SDD detector in 90° geometry. For the presented measurements, the Mo K characteristic radiation of the X-ray tube at 17.4 keV was monochromatized and focused on the sample by a parabolic graded ML optic with variable spacing along the beam axis. The resulting beam size is 5.62 mm x 0.1 mm at 44 mm after the optic with a Gaussian shape in the angular axis. The divergence in this axis amounts to FWHM = $0.014° \pm 0.003°$. Both size and divergence are determined experimentally via imaging with a CCD camera which was placed in the beam path.

The samples are pressed against three contact points of the detector housing ensuring a minimized distance between sample and detector of ~2.73 mm. For angular discrimination, the whole sample-detector assembly is tilted relative to the excitation beam in defined steps.

Typically, two scans per sample were conducted, a long scan with a step size of 0.0025° and measurement times of 30 s per angle spanning about 3° as overview scan and then a more detailed scan around the interference structure with a step size of 0.001° with measurement times of 15 s to 20 s. Both step sizes were well below the determined beam divergence with respect to the incident angle. The measurement time of the complete GIXRF scans amounted to 75 min to 600 min.

Fluorescence intensities were derived through the deconvolution software of the manufacturer (Spectra 7.8.2.0). As shown in a previous work [17], measurements of multilayers with bilayer thicknesses of several nanometers are feasible with the setup. However, the here investigated CrSc multilayers have bilayer thicknesses below 2 nm and suffer from strong intermixing of the individual layers.

To investigate the capabilities of the commercial setup and gain reliable information about the in-depth composition of the samples, one CrSc multilayer (MP15004) was additionally measured with synchrotron radiation. The experiments were carried out in the PTB laboratory at the electron storage ring BESSY II, employing the four-crystal monochromator beamline for bending magnet radiation [18]. PTB's in-house built instrumentation [19] offers the possibility for reference-free XRF experiments as the setup is thoroughly calibrated and the emission of the source fully quantified. While the reference-free approach does not necessarily lower the uncertainties of analysis, uncertainty budget calculations are independent on standards or references which are mandatory for common instrumentation. The setup is installed in an ultra-high vacuum chamber equipped with a 9-axis manipulator, allowing for a very precise sample alignment with respect to all relevant degrees of freedom. The emitted fluorescence radiation is detected by means of a calibrated silicon drift detector (SDD) with an effective active area of 15.8 mm² mounted at 90° with respect to the incident beam with a distance between sample and detector of ~132 mm. This large distance was chosen to minimize the dead time and pile-up effects. Additional calibrated photodiodes on a separate 2θ axis allow for the determination of the incident photon flux.

An incident photon energy of 6.5 keV was chosen which allows to excite both Cr-K and Sc-K shell fluorescence radiation. The synchrotron data was taken at the FCM beamline, where the incident beam is approximately 300 µm x 300 µm in size (FWHM). The flux was about $10^9$ ph/s and the SDD integration time 5 s to 10 s depending on the measurement and incidence angle. For the presented scan of the resonance, we used an angular step size of 0.001°. At each incidence angle, the recorded fluorescence spectra are deconvolved using the known detector response functions [20] for the relevant fluorescence lines as well as physically modelled background contributions. Subsequently, the fluorescence intensities are normalized to detector efficiency, incident photon flux and live time and corrected for the solid angle of detection.

| Sample | Structure (from wafer) | Thickness Cr / nm | Thickness Sc / nm | Thickness B4C / nm | Designed period / nm | XRR measured period / nm |
|---|---|---|---|---|---|---|
| MP15 004 | Cr/B$_4$C/Sc | 0.6 | 1.0 | 0.3 | 1.9 | 1.65 |
| MP15 007 | Cr/B$_4$C/Sc | 0.6 | 1.0 | 0.6 | 2.2 | 1.72 |
| MP15 008 | B$_4$C/Cr/Sc | 0.6 | 1.0 | 0.6 | 2.2 | 1.72 |
| MP15 009 | B$_4$C/Cr/Sc | 0.6 | 1.0 | 0.9 | 2.5 | 1.83 |

Table 1: Design parameters of the investigated chromium scandium multilayer samples. The last column lists the period thickness obtained from X-ray reflectivity measurements [7].

### 3. Samples

Four multilayered samples were available with nominal layer thicknesses and compositions as listed in Table 1 (and shown in Figure 4). The substrate of the samples is 1 mm pure silicon with the multilayer structure magnetron sputtered on top, see [7] for details. In all samples the designed Cr layer is 0.6 nm thick, Sc 1.0 nm, the period is repeated 100 times with a 2.5 nm thick B$_4$C capping layer. For the following analysis this capping layer is omitted, as both elements were not detected and the absorption of this first layer for Cr and Sc K fluorescence is negligible. The samples differ in their sequence of layers and the B$_4$C layer thickness.

Prior to the presented measurements, all samples were analyzed with hard X-ray reflectometry (XRR) [7] with the results displayed in the last column of Table 1. The deviation of the designed to the measured period thickness already suggests a severe intermixing of layers, as already reported for example for W/Si [21] or Ni/SiC [22] ML structures.

### 4. Evaluation

Figure 1 displays as an example energy-dispersive spectra collected with both setups at an incidence angle about 0.1° after the interference structures. Clearly, higher counting statistics are discernible for the laboratory setup. This is on the one hand due to the larger footprint of the excitation on the sample (roughly factor 10) and on the other hand due to the optimized high solid angle of detection of the laboratory setup (chip area: factor 2; distance: factor 50). Besides Si, Cr and Sc, the Ar K fluorescence lines can be discerned, which do not show any angular dependency. Therefore, angular profiles of the Cr K, Sc K and Si K fluorescence of the GIXRF scans were prepared. All scans showed interference structures in the profiles caused by the XSW effect which are located at ~1.25° for the laboratory measurement and 3.35° for the synchrotron measurement.

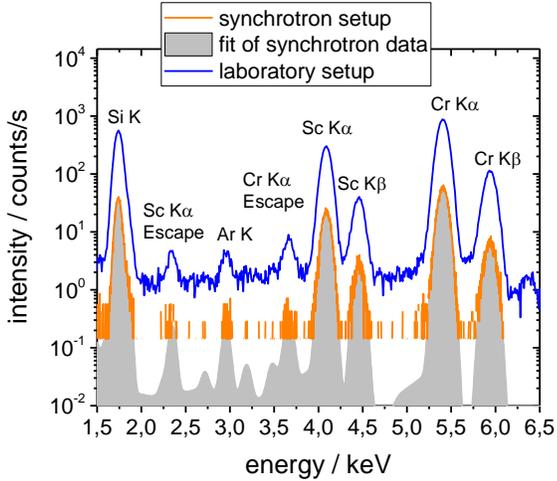

Figure 1: Energy-dispersive spectra taken at 0.1° after the interference structure with both setups. Displayed is also the fit used for the deconvolution of the synchrotron data.

For the laboratory measurements, a Monte Carlo code was developed to calculate the solid angle of detection dependent on the footprint of the excitation beam on the sample and the detection geometry.

To simulate the GIXRF profiles an in-house, C++ based code is employed, which has already been used for a variety of other spectrometers [14, 17, 23-26]. The code forward calculates from a sample model the angular dependent fluorescence intensities based on the solution of the Sherman equation for each relevant fluorescence line taking into account the beam divergence of the incident radiation. The necessary fundamental parameters are derived from the Elam data base [27], with the exception of sub-shell ionization cross sections taken from the Ebel data base [28]. An investigation of secondary effects was performed resulting in negligible contributions, see SI for details, therefore, these were not included in the simulations. Additionally, the algorithm is extended as introduced in de Boer [10] to account for refraction and reflection.

The evaluation consists of two steps: In the first step, the sample model is derived empirically through forward calculations based on initial guesses. In the second step, this model system is optimized with the help of an iterative fitting procedure. Both steps are descried in the following.

For the forward calculation, the sample model must include the number and succession of layers. For each layer the composition, density as well as optical parameters are necessary. The analytical challenge is to establish initial parameters for this calculation and determine the minimal possible number of variable fitting parameters for step 2. The samples were modelled with identical layers in the 100 periods. It must be noted that the information depth of the experiments is limited by the self-attenuation behavior of the excitation and fluorescence radiation. Thus, deeper regions of the multilayer contribute less to the overall observed signals. A severe change of layer geometry as a function of period number can be excluded, because of the measured interference structure in the profiles, but variations in the low percent range are possible and cannot be analyzed with the presented method.

The intermixing of layers can be modelled with additional intermixed layers or for example through roughness introduced with Debye-Waller factors (DWF) at the interfaces. Both approaches were tested and it was found, that for severe intermixing between sub-nanometer layers as in the presented case, fitting without DWF decreases the fitting uncertainties while resulting in equal fit results, see SI for details. Therefore, the approach with additional layers was selected.

| Layer | Density / g/cm$^3$ | Composition / at% |
|---|---|---|
| **CrSc** | 5.7 | Sc60Cr40 |
| **Sc** | 2.81 | Sc100 |
| **CrScB$_4$C** | 4.8 | Sc80Cr10(B$_4$C)10 |
| **CrB$_4$C** | 4.65 | Cr50(B$_4$C)50 |
| **Cr** | 7.05 | Cr100 |

Table 2: Density and composition assumptions for the materials used in the fitting routine.

Step 1:
As a starting point the layer sequence derived from the manufacturing process is assumed. Concerning densities, 2.52 g/cm$^3$ for the B$_4$C layer is defined and for the Sc and Cr layers values as calculated by Haase *et al.* [9] are used (2.81 g/cm$^3$ and 7.05 g/cm$^3$, respectively) due to similar layer thicknesses and production processes. The optical parameters for the layers are calculated from linear combinations of single components extracted from the NIST database [30]. With this model, a forward calculation is performed and compared to the measured data. Subsequently, additional intermixing layers are added between the three layers with fixed elemental ratios. For these mixed layers, the densities and optical constants are calculated from linear combinations according to the assumed composition. Table 2 lists the derived used layer compositions and densities. After each change, the forward calculation is performed and the shape of the GIXRF curves compared to the experimental data until a suitable initial model is found.

For sample MP15004 this procedure resulted in a sequence of CrSc/Sc/CrScB$_4$C/CrB$_4$C/Cr. It was found that no pure B$_4$C layer can be discerned, and an intermixing of CrB$_4$C with Sc rather than B$_4$C with Sc was found to provide better results for the initial sample model. A comparable asymmetric behavior was also found by Haase *et al.* [9]. Thus, this composition model was also used for the other samples.

Step 2:
With this established model, the thicknesses of each individual layer are the only variable fitting parameters in the algorithm with the minimum thickness of 0 nm. The sum of all layer thicknesses and thus the period is restricted by the period thickness as derived for the multilayer from XRR measurements [7], see Table 1, resulting in 4 independent fit parameters (5 layer thicknesses minus period thickness).

## 5. Results

To compare performance and confirm the methodology for the laboratory data the sample MP15004 was measured with both setups - in the laboratory and at BESSY II. The same initial parameters were used to fit the GIXRF data of both data sets. The used algorithm differentiates solely in the deviating instrumental parameters between the two setups: the initial X-ray energy, the beam divergence and the solid angle of detection. In the laboratory setup, a full quantification is not feasible due to the lack of information about the absolute energy distribution of the incident radiation. Therefore, both calculations are normalized to an average intensity value for better comparison. Thus, the shape of the profiles which yields the sample geometry and the layer thicknesses and not the absolute intensities are fitted in the analysis. A full quantification is feasible after calibration of the setup for future analysis. The simulation and fitting time together on an 8.0 GB RAM with an Intel Core i5-3550 CPU with 3.30 GHz machine resulted in a minimal value of 8 hours for one scan, with higher values depending on the number of angular positions.

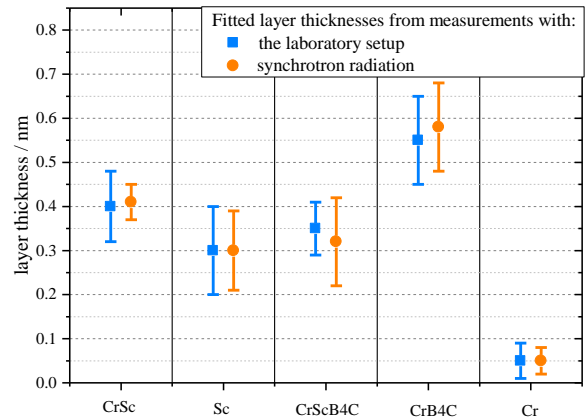

Figure 2: Results of the fitting procedure for data obtained from sample MP15004 with the laboratory setup and the setup at BESSY II. The shown error values are obtained from the square root of the diagonal elements of the covariance matrix of the fitting procedure.

The fitted layer thicknesses are shown in Figure 2 together with calculated fitting error values obtained from the square root of the diagonal elements of the covariance matrix of the fitting procedure. Here, it should be noted that the counting statistics of the synchrotron experiment are lower due to a shorter available measurement time, the smaller footprint on the sample and a significantly smaller solid angle of detection. For all layers, the fitted thicknesses are within the statistical fitting errors of the two data sets. The values of the layer thicknesses from the laboratory fit are additionally listed in Table 3. The good agreement of the values obtained from both data sets indicates that the solid angle simulation and the adjustment strategy in the laboratory spectrometer are suitable for the analysis as also already shown in [17] and that systematic errors are small in respect to the uncertainties derived from the simulation. This is true the more so as the fitting error values underestimate the actual confidence limits as uncertainties in the spectrometer setup are not included (introduced through sample alignment, rotation axis, divergence, deconvolution …).

A comparison of the relevant angular range of the measured GIXRF profiles from the laboratory and synchrotron radiation measurements with the simulated profiles according to the fitted sample models is shown in Figure 3. For ease of comparison, the profiles show the interference features in the Cr and Sc Kα profiles as a function of scattering vector q, with q = 4π/λ sin(θ) where λ is the excitation wavelength and θ the incidence angle. The laboratory data shows superior counting statistics as already mentioned. The difference in the two sample models derived by the two experiments is so small that when using the values derived from one measurement and simulating it for the other measurement, the shape of the GIXRF profile does not change significantly enough to be visibly discriminated.

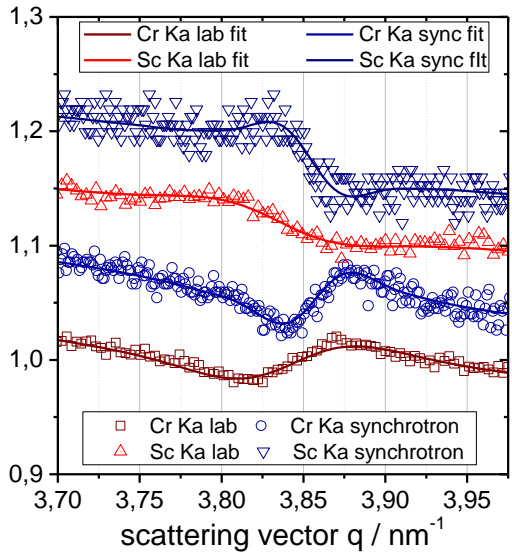

Figure 3: Measured (dots) and fitted (lines) GIXRF data obtained with the two setups for the Cr and Sc Kα fluorescence line of sample MP15004. The graphs are plotted with an offset of 0.06 for clarity.

Thus, the remaining 3 samples, for which no synchrotron data were available, were modeled solely using the measured laboratory data. Other than the order of the individual layers the same routine for fitting was used for the samples as for MP15004 with initial parameters listed in Table 2 and 3. For all samples, the algorithm converged, and the calculated layer thicknesses including the estimated fitting errors are listed in Table 3. For none of the samples, a distinct chromium layer is found, and the chromium is contained solely in the intermixing layers. This is visualized in Figure 4, where the designed and the corresponding fitted period is presented for all samples by stacked box plots.

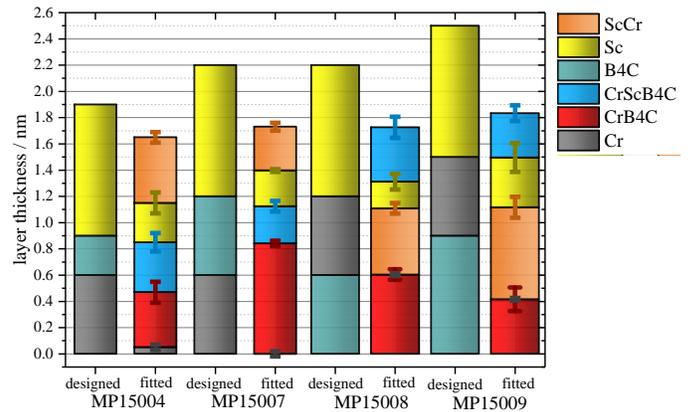

Figure 4: Designed and fitted layer sequence, composition and thickness for the 4 measured samples. The shown error values are obtained from the square root of the diagonal elements of the covariance matrix of the fitting procedure.

| Layer | MP15004 thickness / nm | | MP15007 thickness / nm | | MP15008 thickness / nm | | MP15009 thickness / nm | |
|---|---|---|---|---|---|---|---|---|
| | initial | fitted | initial | fitted | initial | fitted | Initial | fitted |
| CrSc | 0.12 | 0.40 ± 0.04 | 0.55 | 0.33 ± 0.03 | 0.3 | 0.51 ± 0.04 | 0.5 | 0.71 ± 0.08 |
| Sc | 0.5 | 0.30 ± 0.09 | 0.3 | 0.27 ± 0.01 | 0.4 | 0.21 ± 0.06 | 0.2 | 0.37 ± 0.11 |
| CrScB4C | 0.22 | 0.35 ± 0.10 | 0.38 | 0.28 ± 0.04 | 0.27 | 0.41 ± 0.08 | 0.48 | 0.34 ± 0.06 |
| CrB4C | 0.6 | 0.55 ± 0.10 | 0.44 | 0.84 ± 0.02 | 0.6 | 0.61 ± 0.04 | 0.6 | 0.42 ± 0.09 |
| Cr | 0.2 | 0.05 ± 0.03 | 0.05 | 0.00 ± 0.02 | 0.15 | 0.00 ± 0.01 | 0.05 | 0.0 ± 0.01 |

Table 3: Results of the thickness values for the individual layers of the four investigated samples with initial fitting values. The shown error values are obtained from the square root of the diagonal elements of the covariance matrix of the fitting procedure.

In all samples, the measured period thickness is lower than the designed one indicating that the strong intermixing increases the density of the initial layers. No clear trend for the influence of the sputtering order or the thickness of the B$_4$C layer can be observed but for safely drawing such conclusions, a larger number of samples should be measured. It is however observable that solely in the sample MP15004 with the thinnest B4C layer, remnants of a chromium layer are found. The optimal solutions of all other samples resulted in samples without a pure chromium layer.

## 6. Conclusions

In this work a commercial laboratory GIXRF spectrometer is utilized for the analysis of multilayered samples with sub-nanometer single layer thicknesses and strong intermixing. The small thicknesses in combination with the intermixing pose challenges to the analytical interpretation of the data, rendering this investigation a showcase of the reachable performance of measurements in the laboratory. A comparison to synchrotron data exemplifies that such measurements are feasible even with equal or superior counting statistics in the range of minutes to hours.

The feasibility of such measurements with lab equipment renders monitoring and analysis possible on-site at the manufacturers. While for the full quantification of a completely new model system additional methods such as XRR are advisable and computation time is high, such commercial systems are well suited for process control. By monitoring the angular position and height of the interference structure valuable information is gathered without the need for beamtime application at large scale facilities and the accompanying waiting time. And eventually even commercial systems, which are capable of performing both XRR and GIXRF to take advantage of the complementary nature of both techniques, are possible as working demonstrators already exist [31, 32].

Sample-wise, the presented results strongly emphasize the difficulties of manufacturing and in the end employing CrSc multilayers as optics for water window radiation. The functionality of these multilayers crucially depends on the quality and stability of the interfaces between the individual layers within the sample which are disrupted by the intermixing processes. In the presented case of samples with sub-nanometer layer thicknesses, the addition of a B$_4$C layer does not inhibit intermixing substantially. The presented mathematical sample models might not be a full absolute representation of the samples; they however unequivocally demonstrate the intermixing and yield consistent and reproducible results.

GIXRF as stand-alone technique does not necessarily yield a unique solution. Other investigations have shown that for the full characterization of this type of samples a multitude of investigations is necessary [9, 33]. Therefore, for this kind of analysis, good initial models are mandatory which must be defined for a certain class of samples. With this at hand, the possibility to perform quantification of multilayer samples with sub-nanometer layer thicknesses with strong intermixing in laboratories has great potential to support the investigation of novel multilayers towards the ultimate goal of high reflectivity. In fact, as discussed, this work paves the way for short feedback loops and process control, facilitating the development of novel structures and nanomaterials in an unprecedented way.

## 7. Acknowledgements

Parts of this research was performed within the EMPIR project Adlab-XMet. The financial support of the EMPIR program is gratefully acknowledged. It is jointly funded by the European Metrology Program for Innovation and Research (EMPIR) and participating countries within the European Association of National Metrology Institutes (EURAMET) and the European Union.
The authors want to thank the team of BRUKER Nano GmbH who provided us with the possibility to carry out the measurements and offered continuous support in the evaluation.